\documentclass[prl,aps,twocolumn,showpacs]{revtex4}

\usepackage[dvips]{graphicx}
\usepackage{dcolumn}

\newcommand{\be}{\begin{equation}}
\newcommand{\ee}{\end{equation}}
\newcommand{\bea}{\begin{eqnarray}}
\newcommand{\eea}{\end{eqnarray}}

\newcommand{\ds}{\displaystyle}
\begin{document}
\topmargin=-20mm

\title{Local Geometry of the Fermi Surface and Magnetoacoustic Responce of 
Two-Dimensional Electron Systems in Strong Magnetic Fields}

\author{Natalya A. Zimbovskaya}

\affiliation{Department of Physics and Astronomy, St. Cloud State 
University, 720 Fourth Avenue South, St. Cloud MN, 56301 }

\date{\today}

 \begin{abstract}
 A semiclassical theory for magnetotrasport in a quantum Hall system near
filling factor $\nu = 1/2 $ based on the Composite Fermions physical picture 
is used to analyze the effect of local flattening of the Composite Fermion 
Fermi surface (CF-FS) upon magnetoacoustic oscllations. We report on 
calculations of the velocity shift and attenuation of a surface acoustic 
wave (SAW) which travels above the two-dimensional electron system, and we 
show that local geometry of the CF-FS could give rise to noticeable changes 
in the magnitude and phase of the oscillations. We predict these changes to 
be revealed in experiments, and to be used in further studies of the shape 
and symmetries of the CF-FS. Main conclusions reported here could be applied 
to analyze magnetotransport in quantum Hall systems at higher filling factors 
$ \nu = 3/2, 5/2 $ provided the Fermi-liquid-like state of the system.

  \end{abstract}
\vspace{1mm} 

\pacs{71.10.pm, 72.55.+s, 73.43.-f}

\maketitle

\section{Introduction and Background}
\label{sec1}

A two-dimensional electron gas (2DEG) in a strong magnetic field reveals
 rich and complex physics. Near half filling of the lowest Landau level 
($ \nu = 1/2$) the ground state of such a system is shown to be a compressible
 Fermi-liquid-like state of Composite Fermions (CF) \cite{one}. These 
quasiparticles are 
distributed inside the  Composite Fermion Fermi Surface (CF-FS). A similar 
physical picture could be adopted to describe the 2DEG at half filling of the
 next Landau level $(\nu = 3/2).$ Experimental evidences of the CF Fermi sea 
at $ \nu = 1/2,3/2$ were repeatedly obtained during the recent decade 
\cite{two}.

 For higher  filling factors close to ${\nu} =N+1/2$ where 
$N \geq 3$ is an integer, the exchange interaction would lead to an 
instability towards charge density wave formation in the relevant Landau
levels. The ground state of the 2DEG for these filling factors  corresponds
to a charge density wave (CDW) and  has a striped structure 
\cite{three,four,five,six}. It could be described as a sequence of one 
dimensional stripes alternating between the adjacent filling factors N and 
N + 1. This  gives rise to a strikingly anisotropic transport properties of 
the 2DEG at half filling of higher Landau levels \cite{seven,eight} which were 
revealed in experiments \cite{nine,ten,eleven}.

 The quantum Hall state at $ \nu = 5/2 $ is perhaps the most enigmatic 
due to its position in the magnetic field spectrum between the high Landau 
level $ N \geq 3$ stripe phases and the low Landau level $ (N \leq 1)$ 
Fermi-liquid like states. Theoretical studies of this state started from the
 model of paired CFs \cite{twelve}. Depending on the interaction strength 
within the system, the 2DEG at $ \nu = 5/2 $ could reveal a striped state, a 
Fermi--liquid, or the paired state \cite{thirteen}. Numerical simulations 
presented in \cite{thirteen} give grounds to believe that at $ \nu = 5/2$  
the CF Fermi liquid undergoes condensation to the paired state at low 
temperature limit. Also, there could be a transition from  the Fermi--liquid 
to the striped phase when in-plane magnetic field is applied. Recently, an 
experimental evidence of the CF-FS at $ \nu = 5/2$ was obtained 
\cite{fourteen}.

So, both theory and experiment give grounds to believe that physical picture 
of CFs which form a Fermi sea  could be 
succesfully employed to describe the 2DEG states at half filling of the 
lowest Landau levels $(N \leq 2$). However, the geometry of the CF-FS was not 
analyzed up to present. It is usually assumed that the CF Fermi liquid is  
isotropic, and the CF-FS is a circle in the two-dimensional quasimomenta 
space. This is an obvious oversimplification. Real samples commonly used in 
studies of the quantum Hall effect have 2DEGs deposited in GaAs/AlGaAs 
heterostructures. Therefore the crystalline field of the host semiconductor 
could significantly influence the CF-FS geometry distorting the original Fermi 
circle \cite{fifteen}. Another sourse of the CF-FS anisotropy, especially 
for higher filling 
factors $ (\nu =3/2,5/2)$, is the interaction in the electron system. 
The development of highly anisotropic charge-density wave formations (striped 
phases) at high filling factors including $5/2,$ gives us strong arguments to 
expect these interactions to work as an extra crystalline  field at the 
Fermi-liquid state of the 2DEG at $ \nu = 3/2,5/2$. As a result the CF-FS 
shape could be further modified.

Theory of magnetotransport in metals showes that the FS local geometry
noticeably affects the electron response of the metal to an external 
perturbation \cite{sixteen}. The change in the response occurs under the 
nonlocal regime of propagation of the disturbance when the mean free path of
 electrons $ l $ is large  compared to the wavelength of the disturbance 
$ \lambda. $ The reason is that in this nonlocal regime only those electrons 
whose motion is somehow consistent with the propagating perturbation can
 strongly absorb its energy. These ''efficient'' electrons are concentrated on
 small ''effective'' segments of the FS.

When the FS includes flattened segments  it leads to an enhancement of
the contribution from  these  segments to the electron
density of states (DOS) on the FS. Usually this enhanced contribution is
small compared to the main term of the DOS which originates from all the
remaining parts of the FS. Therefore it cannot produce noticeable changes
in the response of the metal under the local regime of propagation of the
disturbance $ (l << \lambda) $ when all segments of the FS contribute to
the response functions essentially equally. However the contribution to the
DOS from the flattened section can be congruent to
the contribution of a small ''effective'' segment of the FS.  In other
words when the curvature of the FS becomes zero at some points on an
''effective'' part of the FS it can give a significant enhancement of
efficient electrons and, in consequence, a noticeable change in the
response of the metal to the disturbance.

Due to the same reasons we can expect local geometrical features of the CF-FS 
to give significant effects on the 2DEG response to an external disturbance.
 As well as for convenient metals, these effects are to be revealed within a 
nonlocal regime $ (l > \lambda).$  It was shown before that the local 
flattening of the CF-FS could give rise to a strong anisotropy in the response
 of a 2DEG to a surface acoustic wave (SAW) \cite{seventeen}. Such 
anomaly was observed in experiments on modulated 2DEG near $ \nu = 1/2$ 
\cite{eighteen}.

Here, we analyze the influence of the CF-FS local geometry on so called 
geometric resonances which were repeatedly observed in 2DEGs in strong 
magnetic field \cite{two}, as well as in convenient metals \cite{nineteen}. 
These oscillations could be revealed within a nonlocal regime, and they 
appear due to periodical reproduction of the most favorable conditions for 
the resonance absorption of the energy of the external disturbance by 
quasiparticles at stationary points on the cyclotron orbit where they move  
along the wave front of the disturbance. When the external disturbance is 
associated with an acoustic wave these geometric resonances are also called
 magnetoacoustic oscillations \cite{nineteen}.
In the following analysis we mostly consider magnetoacoustic oscillations in 
the  2DEG at $ \nu = 1/2$ state, and we describe this state within the 
framework of Halperin, Lee and Read (HLR) theory \cite{one}. However, we 
believe that the main results of the present analysis could be applied to 
study magnetotransport in 2DEGs at higher filling factors $ (3/2,5/2)$ 
provided that the system is at a Fermi-liquid-like state.

\section{Main Equations}
\label{sec2}

Due to the piezoelectric properties of GaAS, the velocity shift
$ (\Delta s/s) $ and the attenuation rate $ (\Gamma) $ for the SAW
propagating along the $ x $ axis across the surface of a
heterostructure containing 2DEG, take the form \cite{tventy}:

                    \begin{equation}
\frac{\Delta s}{s} = \frac{\alpha^2}{2} \mbox{Re}
\left( 1 + \frac{i \sigma_{xx}}{\sigma_m} \right )^{-1} ;
                    \end{equation}

                    \begin{equation}
\Gamma = -q \frac{\alpha^2}{2} \mbox{Im} \left( 1 + \frac{i
\sigma_{xx}}{\sigma_m} \right )^{-1} .
                   \end{equation}
 Here $ {\bf q},\omega = s q $ are the SAW wave vector and frequency, 
respectively, $ \alpha $ is the piezoelectric coupling constant, $ \sigma_m =
\varepsilon s/2 \pi, \; \varepsilon $ is an effective dielectric
constant of the background and $ \sigma_{xx} $ is the component of
the electron conductivity tensor.

According to HLR theory, the electron resistivity
tensor $ \rho $ at $ \nu = 1/2 $ is given by:

                \begin{equation}
                \rho = \sigma^{-1} = \rho^{CF}  + \rho^{CS}
                                  \end{equation}
where $ \rho^{CF} $ is the CF resistivity tensor,
and the contribution $\rho^{CS}$  originates from the Chern--Simons 
formulation of the theory. This tensor contains only off diagonal 
elements $ \rho_{xy}^{CS} = - \rho_{yx}^{CS} = 4 \pi \hbar /e^2. $

The CF resistivity tensor $\rho^{CF}$ could be calculated as the inverse for 
the CF conductivity $ \tilde \sigma \ (\rho^{CF} = \tilde \sigma^{-1}).$ We 
carry out our analysis in a regime where $ \rho_{xx} \rho_{yy} <<\rho_{xy}^2,$
 therefore the relevant component of the electron conductivity could be 
written in the form:
 
 \be
 \sigma_{xx} (q) = \frac{e^4}{(4 \pi \hbar)^2} \rho_{yy}^{CF} = 
\frac{e^4}{(4 \pi \hbar)^2} \frac{\tilde \sigma_{xx}}{\tilde \sigma_{xx} 
\tilde\sigma_{yy} + (\tilde \sigma_{xy})^2} .
 \ee
 The CFs are supposed to experience not actual but 
reduced magnetic field $ B_{eff} = B - B_{1/2}$ where $ B_{1/2} $ corresponds 
to one half filling of the lowest Landau level. Their motion could be 
described within a semiclassical approximation based on the Boltzmann's 
transport equation. Following standard methods \cite{tventyone}  we obtain:
 
  \be
\tilde \sigma_{\alpha \beta} = \frac{m^*e^2}{2 \pi \hbar^2} \sum_n
 \frac{v_{n \beta} (-q) v_{n\alpha} (q)}{i n \Omega - i \omega + 1/\tau}.
  \ee
  Here, $m^*, \Omega $ are the CF cyclotron mass and their cyclotron 
frequency at the field $ B_{eff}; \ \tau $ is the CF scattering time, and 
$ v_{n \alpha} (q)$ are the Fourier transforms of the CF velocity components:

 \be
 v_{nx} (q) = \frac{n}{2 \pi} \frac{\Omega}{q} \int_0^{2\pi} d \psi 
 $$$$ \times \exp 
\left \{ in \psi - \frac{i q}{\Omega} \int_0^\psi v_x (\psi') d \psi'\right \};
 \ee

 \be
 v_{ny} (q) = \frac{1}{2 \pi} \int_0^{2\pi } d \psi v_y (\psi)
 $$$$ \times \exp 
\left \{ in \psi - \frac{i q}{\Omega} \int_0^\psi v_x (\psi') d \psi'\right \}.
 \ee
 The variable $ \psi $ included in these expressions is the angular coordinate
 of the CF cyclotron orbit.

The most favorable conditions for magnetoacoustic oscillations to be revealed 
occur at moderately strong effective magnetic field when $ ql >> \Omega 
\tau >> 1.$ Under these conditions the main contributions to the integrals 
over $ \psi $ in the expressions (6),(7) come from the neighborhoods of 
stationary points at CF cyclotron orbits. So, the expressions for
 $ v_{n\alpha} (q)$ could be rewritten as:

  \be
 v_{nx} (q) = \frac{n \Omega}{q} \cos \left ( qR - \frac{\pi n}{2} - \Phi
\right ) X (qR) ;
 \ee

  \be
 v_{ny} (q) = -i \sin \left ( qR - \frac{\pi n}{2} - \Phi \right ) V(qR) .
 \ee 
 Here, $ 2 R $ is the diameter of the CF cyclotron orbit in the direction 
of propagation of the SAW, $ X (qR)$ is a dimensionless quantity, and $ V(qR)$
has dimensions of velocity.  Both  $ X (qR)$ and $ V(qR)$ are power functions 
of $ qR$ with the same exponent. As we show below, the value of the exponent 
is determined with the local geometry of small effective segments of the 
CF-FS which correspond to the vicinities of the stationary points. When these 
segments are 
flattened, this leads to  significant changes in the magnitude of the 
magnetoacoustic oscillations.

\section{The CF-FS model}
\label{sec3}

Within the commonly used  isotropic model of CF Fermi liquid at $ \nu = 1/2 $ 
the CF-FS is a circle, and its radius  $ p_F$ equals $ \sqrt{4 \pi N\hbar^2}$ 
where "N" is the electron density. To develop more realistic model of the 
CF-FS we include a periodic static electric field applied along the "y" 
direction which provides CFs with the potential energy of magnitude $ U_g \ 
(\bf g$ is the wave vector of the electric field). The above electric field 
could originate from  interactions with electrons of lower Landau 
levels at $ \nu = 3/2, 5/2$ and from the crystalline field of the host 
semiconductor. The latter is especially important at  $ \nu = 1/2. $ The 
point is that wherever it comes from, this field distorts the CF-FS, 
including formation of local anomalies of the CF-FS curvature. 

Assume for simplicity that the electric modulation is weak $(U_g << E_F,$ 
where $ E_F$ is the CFs Fermi energy). Then we can use the 
nearly-free-electron model to derive the energy-momentum relation for the 
CFs. When the modulation period is small enough $ \hbar g > 2 p_F$ we obtain:

    \begin{equation}
E({\bf p}) = \frac{p_x^2}{2 m} +  \frac{p_y^{*2}}{2 m} +
\frac{(\hbar g)^2}{8 m} -
\sqrt {\left (\frac{\hbar g p_y^*}{2 m} \right)^2 + U_g^2} \, .
                   \end{equation}
Here $ p_y^* = p_y - \hbar g/2, \;m $ is the CF effective mass;
$ U_g $ is the magnitude of the quasiparticle potential energy
in the periodic electric field.  Calculating the FS curvature: 

  \be
 K = \frac{1}{v^3}\left ( 2 v_x v_y \frac{\partial v_x}{\partial p_y} -
v_x^2 \frac{\partial v_y}{\partial p_y} - v_y^2 
\frac{\partial v_x}{\partial p_x} \right)
      \ee
 with $ v = \sqrt{v_x^2 + v_y^2} $ one can find it tending to zero when
$p_x $ tends to $ \pm p_F (U_g/E_F)^{1/2} $ (See Fig.1).

\begin{figure}[t]
\begin{center}
\includegraphics[width=6.2cm,height=6.7cm]{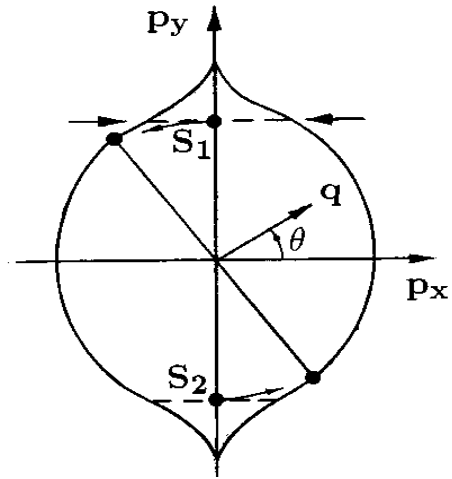}
\caption{
The shape of the CF-FS in the nearly-free-electron approximation (solid line),
and the CF-FS described with the Eq. (14) (dashed line). Point $ S_1$ and 
$ S_2$ are associated with the stationary points at the CF cyclotron orbit 
when the SAW wave vector $ \bf q $ points in the $ p_x $ direction.
}  
\label{rateI}
\end{center}
\end{figure}

In the vicinities of the corresponding points on the FS the
quasiparticles velocities are nearly   parallel to the  the $y$ direction. 
Near these zero curvature points we can derive an asymptotic expression for 
energy-momentum relation (10). Introducing
$(p_{x_0}, p_{y_0})$ by $ p_{x_0} = \zeta p_{F} $, $ \displaystyle
{p_{y_0} = p_{F} \left (1-\frac{1}{\sqrt{2}}\zeta^{2} \right )} $, where
$\zeta=\sqrt{U_{g}/E_{F}}$, we can expand the
variable $p_{y}$ in powers of $(p_{x} - p_{x_0}), $ and keep the lowest
order terms in the expansion. We obtain:

       \begin{equation}
       p_{y}-p_{y_0} = -\zeta(p_{x} - p_{x_0}) -
       \frac{2}{\zeta^{4}}\frac{(p_{x} - p_{x_0})^{3}}{p_{F}^{2}}.
                    \end{equation}
Near $p_{x0}$, where $(|p_{x} - p_{x_0}| < \zeta^{2}p_{F}) $ the first
term on the right side of Eq.(12) is small compared to the second one and
can be omitted. So we have:

\begin{equation}
E({\bf p})=\frac{4}{\zeta^{4}}\frac{p_{F}^{2}}{2m}
\left(\frac{p_{x}-p_{x_0}}{p_{F}}\right)^{3}+\frac{p_{y}^{2}}{2m}.
          \end{equation}
The ''nearly free'' particle model can be used when $ \zeta^2 $ is very
small. For larger $ U_g $ the local flattening of the CF-FS can be
more significant. To analyze the contribution to the conductivity from these flattened parts we  generalize the expression for $E({\bf p})$ and define our
dispersion as:

 \begin{equation}
 E({\bf p}) = \frac{p_{0}^{2}}{2m_{1}} \left |
\frac{p_{x}}{p_{0}} \right |^{\gamma} + \frac{p_{y}^{2}}{2m_{2}},
             \end{equation}
 where $p_{0}$ is a constant with the dimensions of momentum,  
$m_{i}$ are the  effective masses, and $ \gamma  $ is a dimensionless
parameter which determines the shape of the CF--FS .
When $\gamma> 2 $ the  CF--FS looks like an ellipse
flattened near the vertices $(0, \pm \sqrt{m_2/m_1} p_{0})$.
Near these points the curvature is:

 \begin{equation}
 K = - \frac{\gamma(\gamma-1)}{2p_0 \sqrt{m_1/m_2}}
\left| \frac{p_x}{p_0} \right|^{\gamma-2}
\end{equation}
 and, $ K \rightarrow 0$ at $p_{x} \rightarrow 0 $. The CF-FS will be
the flatter at $ p_x = 0$,
the larger is the parameter $ \gamma  $. A separate investigation is required
to establish how $ \gamma $ depends on $ U_g.$ Here we postulate Eq.(14) as
a natural generalization of Eq.(13).

 When $ p_F > \hbar g $ we have to consider
the CF--FS as consisting of several branches belonging to several
''bands'' or Brillouin zones. The modulating potential wave  vector {\bf g} 
in this case determines the size of the ''unit cell''.  However
with this condition we also may  expect some branches of the CF--FS to be
flattened. Within an appropriate geometry of an experiment these flattened 
segments of the CF-FS become the effective parts of the FS. Consequently, 
the response of the CF system to the SAW could undergo significant  changes.
Prior to start the analysis of these changes we remark that our model of the 
deformed CF-FS (14) could be easily generalized and accomodated to more 
complicated geometry of electric field which determines the CF-FS shape and 
symmetries. However, even the simple model (14) captures the essential 
physics, enabling to include local flattenings of the CF-FS into 
consideration. Therefore we adopt this model in the further analysis.

\section{Results and Discussion}

When the SAW propagates along the "x" direction the vertices $S_1,S_2$ of the 
flattened ellipse (14) correspond to the stationary points on the CF 
cyclotron orbits (see Fig. 1). The enhanced DOS of quasiparticles in their 
vicinities influences the features of the magnetoacoustic oscillations. 
Using the stationary-phase method we obtain the following asymptotics for 
the functions $ X (qR)$ and  $ V (qR):$ 

 \be 
 X(qR) =  \frac{m^*}{p_0} V(qR) $$$$ =
  \frac{2}{\pi}\left (\frac{m^*}{\sqrt{m_1m_2}} \right )^{1/\gamma}
\frac{\Gamma (1/\gamma)}{ \gamma} 
\left(\frac{2}{qR} \right)^{1/\gamma}.
 \ee
 Here, $\ \ds R = \frac{cp_0}{e B_{eff}} \sqrt{\frac{m_2}{m_1}}, \ $ 
 $ \Gamma (x) $ is the gamma function. As for the phase shift $ \Phi,$ 
it appears to be equal $ \pi /2 \gamma.$

 Using these results (16),  as well as standard formulas \cite{tventytwo}:

\be
 \sum_{n = - \infty}^\infty \frac{1}{\omega + i/\tau - l \Omega} 
= - \frac{i \pi}{\Omega} \coth \frac{\pi (1 - i \omega \tau)}{\Omega \tau} ;
 \ee

\be
 \sum_{n = - \infty}^\infty \frac{(-1)^n}{\omega + i/\tau - l \Omega} 
 $$$$
 = - \frac{i \pi}{\Omega} \frac{1}{\sinh \big[\pi (1 - i \omega \tau) /
\Omega \tau\big ]}
 \ee
 we can transform the expressions (5) for the CF conductivity components to 
the form:

\be
 \tilde \sigma_{xx} = \frac{2 }{\rho_0} \frac{b^2(1 - i \omega \tau)}{(q l)^2};
 \ee

  \be
 \tilde \sigma_{xy} = - \sigma_{yx} = - \frac{2}{\rho_0} 
\frac{g^2}{(ql)^2} \bigg(\frac{qR}{2} \bigg)^{1 - 2/\gamma} 
 $$$$ \times
\frac{(1 - i \omega \tau) \sin \big(2 qR - \pi/\gamma \big)}
{\sinh \big[ \pi (1 - i \omega \tau) / \Omega \tau\big ] } ;
  \ee

\be
 \tilde \sigma_{yy} = \frac{2}{\rho_0} \frac{d^2}{ql} 
\left (\frac{qR}{2}\right)^{1 - 2/\gamma} 
\left \{ \coth \frac{\pi (1 - i \omega \tau)}{\Omega \tau} \right.
 $$$$   \left.
 - \cos \left (2qR - \frac{\pi}{\gamma} \right ) 
\sinh^{-1} \frac{\pi (1 - i \omega \tau)}{\Omega \tau} \right \}.
 \ee
  where $ \rho_0 = m^*/ Ne^2 \tau $ is the CFs Drude resistivity; $ \ds l = 
\frac{\tau}{m^*} \sqrt{\frac{A}{\pi}};$ and $ A$ is the area of the CF-FS. 
The factors $ b^2, d^2 $ and $ g^2$ included in (20)--(22) are the 
dimensionless constants of the order of unity. Expressions for these 
constants are omitted for brevity. 
For a circular CF-Fs we have: $ \gamma = 2,  \ p_0 = p_F, \
 b^2 = g^2 = d^2 = 1,$ and our expressions (19)--(21) take on the form:

 \be
\tilde \sigma_{xx} = \frac{2}{\rho_0} \frac{1 - i \omega \tau}{(ql)^2} ;
 \ee

 \be
\tilde \sigma_{xy} = - \tilde \sigma_{yx} = \frac{2}{\rho_0} 
 \frac{1 - i \omega \tau}{(ql)^2} \cos(2qR) 
 $$$$ \times
 \sinh^{-1} 
\left [\frac{\pi(1-i\omega \tau)}{\Omega \tau} \right ];
  \ee

 \be
 \tilde \sigma_{yy} = \frac{2}{\rho_0} \frac{1}{ql}
\left \{ \coth \left [\frac{\pi(1-i\omega \tau)}{\Omega \tau} \right ]
 \right.   $$$$ \left.
 - \sin (2qR) \sinh^{-1}\left [\frac{\pi(1-i\omega \tau)}{\Omega \tau} 
\right ]  \right \}.
 \ee

Comparison of our expressions (19)--(21) with the results for a Fermi circle 
showes that the local flattening of the CF-FS near the points which 
correspond to the stationary points of the CFs cyclotron orbit  enhances  
the magnitude of the magnetoacoustic oscillations of the CF conductivities. 
 A similar effect was 
studied before for conventional metals \cite{tventythree}. The effect 
originates from the enhancement of the quasiparticle DOS at flattened 
segments of the FS.

The above considered enhancement of magnetoacoustic geometric oscillations 
could be manifested only when the stationary points on the CF cyclotron 
orbit correspond to the points located at flattened segments of the CF-FS. 
Therefore the effect has to be very sensitive to variations in the direction 
of the SAW propagation. Suppose that the SAW travels  at some angle $ \theta$ 
with respect to the symmetry axis of the CF-FS  as it is shown in Fig. 1. 
Then the stationary points slip from the flattened pieces and fall into 
"normal" segments of the FS whose curvature takes on  nonzero 
values. Due to the lower DOS of quasiparticles at these "normal" CF-FS 
segments, the number of efficient CFs which can participate in the absoption 
of the SAW energy decreases when the angle $ \theta $ increases. This results 
in the noticeable reduction of the oscillations.

Assuming a nonzero value for the angle $ \theta, $ we can present Fourier 
transforms of the CF velocity components in the form:

  \be
 v_{nx} (q) = \frac{n \Omega }{q} \left [ 
\cos \left ( qR - \frac{\pi n}{2} \right ) S_\gamma (qR, \theta) \right.
 $$$$ \left.
 +  \sin \left ( qR - \frac{\pi n}{2} \right ) W_\gamma (qR, \theta) \right ];
 \ee 

 \be 
 v_{ny} (q) = - \frac{i p_0}{m^*} \left [ 
\sin \left ( qR - \frac{\pi n}{2} \right ) S_\gamma (qR, \theta) \right.
 $$$$ \left.
  - \cos \left ( qR - \frac{\pi n}{2} \right ) W_\gamma (qR, \theta) \right ].
 \ee
 Here,

 \be
 S_\gamma (qR, \theta) =  \frac{2}{\pi} 
\left (\frac{m^*}{\sqrt{m_1 m_2}} \right )^{1/\gamma}
 $$$$ \times  \int_0^\infty 
 \cos \left [ \frac{qR}{2} \left ( \frac{m_2}{m_1} \sin^2 \theta y^2
+ \cos^{\gamma} \theta y^{\gamma} \right ) \right ];
 \ee

 \be
 W_\gamma (qR, \theta) =  \frac{2}{\pi} 
\left (\frac{m^*}{\sqrt{m_1 m_2}} \right )^{1/\gamma}
 $$$$ \times \int_0^\infty
\sin \left [ \frac{qR}{2} \left ( \frac{m_2}{m_1} \sin^2 \theta y^2
+ \cos^{ \gamma} \theta y^{ \gamma} \right ) \right ].
 \ee
  Now, we expaind these functions $ S_\gamma $ in series in powers of a 
dimensionless parameter $\ \ds \ \xi \  \ \bigg(\xi = \frac{m_2}{m_1} \bigg(
\frac{qR}{2} \bigg )^{1 - 2/\gamma} \\ \times  \tan^2 \theta \bigg).$
For small angles $ \theta $ we have $ \xi << 1,$ and the expansion accepts 
the form \cite{tventytwo}:

 \be
 S_\gamma (qR, \theta) = \frac{1}{\gamma \cos \theta} 
\left (\frac{2}{qR} \right )^{1/\gamma}
  $$$$ \times
 \sum_{r=0}^\infty \frac{(-1)^r}{r!} \xi^r \Gamma 
\left (\frac{2r +1}{\gamma} \right )
\cos \left [ \pi \frac{1 - r (\gamma - 2)}{2 \gamma} \right ].
 \ee
 As $ \theta $ increases to the values guaranteeing the inequality $ \xi > 1$ 
to be valid, we have to use a different power expansion, namely:

 \be
 S_\gamma (qR, \theta) = \frac{1}{2 \sin \theta} 
\sqrt{\frac{m_1}{m_2} \frac{2}{qR}}
 \sum_{r=0}^\infty \frac{(-1)^r}{r!} \xi^{- \gamma r/2}
    $$$$ \times  \Gamma
\left ( \frac{\gamma r + 1}{2} \right ) 
\cos \left [ \frac{\pi}{4} \bigg (r(\gamma - 2) +1\bigg) \right ].
 \ee
 Expansions for the function $ W_\gamma (qR, \theta)$ could be ontained from 
(29),(30) by replacing cosines by sines of the same angles.

When $ \theta \to 0$  we arrive at our former asymptotic expressions for the CF 
velocity components (16). While $ \theta $ increases, the 
functions $ S_\gamma (qR, \theta)$ and $ W_\gamma (qR,\theta)$ diminish and 
approach 
the value $ (\pi m_1/4m_2 qR)^{1/2}$ which is 
a typical estimate for an elliptical CF-FS as $ \theta \to 90^o. $ Angular 
dependence of the amplitude factor of magnetoacoustic oscillations $ F_\gamma^2 
(qR,\theta) =S_\gamma^2 (qR,\theta) + W_\gamma^2 (qR,\theta) $ is 
presented in the Fig. 2. We see in this figure that the effect of local 
flattening of the CF-FS on the oscillations amplitude remains 
distinguishable  even for moderate flattenings $ (\gamma = 4).$

\begin{figure}[t]
\begin{center}
\includegraphics[width=5.0cm,height=7.0cm]{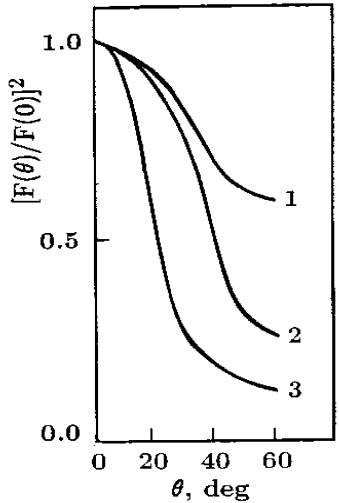}
\caption{
Angular dependence of the amplitude of magnetoacoustic oscillations. The 
curves are plotted for $ \  qR = 10; \ m_1 = m_2; \ \gamma = 4 $ (curve 1);
 $\ \gamma = 6 \ $ (curve 2); and $ \ \gamma = 8 $ (curve 3). 
}
\label{rateI}
\end{center}
\end{figure}
Differences in  magnitudes of geometric osclillations of the CF conductivity 
components are manifested  in the electron conductivity. Substituting  our 
results for $  \tilde \sigma_{\alpha\beta }$ into (4) we have:

\be 
 \sigma_{xx} = \frac{qe^2}{8d^2 p_0}
\left ( \frac{2}{q R} \right )^{1 - 2/\gamma}
  $$$$   \times
 \frac{\sinh [\pi (1 -i \omega \tau)/\Omega \tau]}{\cosh 
[\pi (1 -i \omega \tau)/\Omega \tau] - \cos (2q R - \pi/ \gamma)} .
 \ee
 We compare this expression with the corresponding result for a circular CF-FS

\be 
 \sigma_{xx}^{el} = \frac{qe^2}{8 p_F} 
\frac{\sinh [\pi (1 -i \omega \tau)/\Omega \tau]}{\cosh 
[\pi (1 -i \omega \tau)/\Omega \tau]- \sin(2q R)},
 \ee  
  and we see that both amplitude and phase of the geometric oscillations in 
the   electron conductivivty differ from those for the CF Fermi circle. The 
same could be applied to magnetoacoustic oscillations described with the 
expressions (1),(2). When the 
effective parts of the CF-FS are flattened, the amplitude of the 
magnetoacoustic oscillations drops.
 We also conclude that  varying the direction of propagation of the SAW we can 
observe angular dependence of the oscillations amplitude. The latter 
originates from the angular dependence of the CF conductivities discussed 
before. 
We have grounds to expect it to be revealed in experiments.

Finally, we belive that Fermi-liquid  state of a 2DEG in quantum Hall regime 
at $ \nu = 1/2,3/2,5/2 $ is anisotropic and exhibits an anisotropic CF-FS. The 
CF-FS geometry reflects symmetries of the crystalline fields of the host 
semiconductor.  For higher 
filling factors  $ \nu = 3/2, 5/2$ it also  could show effects of
interactions in the electron system. It was already found that screening 
due to polarization of remote Landau levels plays an essential role for the 
preferred orientation of the stripes induced by an in-plane magnetic field at 
$\nu = 5/2$ \cite{tventyfour}. Therefore, we  may conjecture that 
at weaker in-plane fields when Fermi-liquid like state of the 2DEG still exits, 
these polarization effects could give extra anisotropies to the Cf-FS.

Anisotropic CF-FSs usually include some flattened segments. Even a naive model 
(10) based on the nearly-free-electron approach, demonstrates that local 
flattening of the CF-FS appears as a result of a weak deformation of the 
latter with an electric modulation. In general, local flattenings are 
initiated with electric fields acting within a 2DEG like crystalline fields 
in usual metals. Accordingly, locations of the flattened segments conform 
with the symmetries of the CF-FS and could reveal these symmetries. The 
results of the present analysis show that magnetoacoustic oscillations in the 
velocity shift and attenuation of the SAW travelling in piezoelectric 
GaAs/AlGaAs heterostructures above the 2DEG, could be used as a tool to 
discover local flattenings at the CF-FS when the 2DEG is in the quantum Hall 
regime at Fermi-liquid like state. This could give a new knowledge of the 
shape and symmetries of the CF-FS  and, consequently, a better understanding 
of the magnetotransport in quantum Hall systems near half filling of lowest 
Landau levels. It would be especially interesting to compare symmetries of 
the CF-FS at the Fermi-liquid like state of the 2DEG at $ \nu = 5/2 $ with 
the characteristic symmetries of the striped state of the system at the same 
filling factor. It is possible that such comparison would give some unusual 
results providing a new insight in the nature of transition from the 
Fermi-liquid to the striped phase of the 2D electron system.

\vspace{2mm}

{\bf \it  Acknowledgments:}
The author thanks G.M. Zimbovsky for help with the manuscript.

\end{document}